\documentstyle[aasms4,12pt]{article}

\def\gs{{_>\atop^{\sim}}}
\def\ls{{_<\atop^{\sim}}}
\newcommand{\kms}{\,km~s$^{-1}$}
\newcommand{\lya}{Ly-$\alpha$}

\include{AAS_defs}

\begin{document}



\title{Q1208+1011: Search for the lensing galaxy.}

\author{Aneta Siemiginowska$^1$, Jill Bechtold$^2$, Thomas L. Aldcroft$^1$ \\
K. K. McLeod$^3$, Charles R. Keeton$^1$}

\bigskip\bigskip

\affil{$^1$Harvard-Smithsonian Center for Astrophysics \\
60 Garden Street, MS-70,  Cambridge, MA~02138, USA \\
 asiemiginowska@cfa.harvard.edu\\
 taldcroft@cfa.harvard.edu\\
 ckeeton@cfa.harvard.edu\\
$^2$ Steward Observatory, Tuscon, AZ 85721, USA\\
jbechtold@as.arizona.edu\\
$^3$ Whitin Observatory, Wellesley College, Wellesley, MA 02181, USA\\
kmcleod@wellesley.edu}

\begin{center}
\bf
\today

\end{center}


\begin{abstract}

We present a high-resolution spectrum of the high redshift, lensed
quasar Q1208+1101, obtained with the echellette spectrograph on the
Multiple Mirror Telescope.  We examine the new and published spectra
and provide an updated list of high-confidence metal-line absorption
systems at $z=1.1349, 2.8626, 2.9118, 2.9136, 2.9149$.  Combining this
with a simple model of the gravitational lens system allows us to
constrain the possible lens redshifts.  The high-redshift ($z > 2.5$)
and low-redshift ($z < 0.4$) candidates can be ruled out with high
confidence. The current spectra effectively probe about 40\% of the
redshift range in which the lens is expected.  In that range, there is
only one known metal-line absorption system, an MgII absorber at
$z=1.1349$.  We consider the possibility that this system is the
lensing galaxy and discuss the implied parameters of the galaxy.

\end{abstract}

\keywords{gravitational lens-quasars-Q1208+1101}

\section{Introduction.}

The bright, high redshift (z=3.815) radio quiet quasar Q1208+1011 has
been identified as a gravitational lens by Maoz et al. (1992) and
Magain et al. (1992).  The lens consists of two components (V=18.3 and
19.8 mag, Bahcall et al. 1992) separated by 0.\arcsec 47, with a 4:1
intensity ratio.  The FOS HST spectra (Maoz et al., 1992) show that
both components have the same redshift and similar spectra.

There are three key aspects in studying gravitational lenses: 1)
understanding the lens geometry; 2) understanding the properties of
the lensing galaxy; and 3) understanding the properties of the
background source. Determining the amount of magnification allows us
to understand the intrinsic quasar emission. Given a limiting observed
magnitude, lensing allows us to probe both to lower intrinsic
luminosities (at a certain redshift) or to higher redshifts (at a
certain luminosity).

Q1208+1011 is apparently an extremely high luminosity source with an
observed optical luminosity of $\sim 10^{48}$~ergs~s$^{-1}$.  The true
intrinsic luminosity is likely to be much lower, which affects the
modeling and influences the parameters of the quasar models such as
required black hole mass or accretion rates (Czerny 1994, Antonucci
1994, Siemiginowska et al 1996).

Precise lens modeling and evaluation of the quasar magnification
requires detailed information about the lens, including its exact
position relative to the quasar images, its morphology or mass
distribution, and its redshift (Kochanek 1991). The SIS lens model
predicts an average magnification of about 4 (Turner et al. 1984),
however we cannot give a correct value for Q1208+1011 until the lens
is detected.

Bechtold (1994) investigated the proximity effect in the spectra of
Q1208+1101 and concluded that the data were consistent with
a magnification of 1.  Fontana at al. (1997) give a factor of 20
magnification for Q1208+1101 based on high resolution Lyman alpha
forest data.  Lens detection combined with the proximity effect could
give stronger constraints on the magnification factor, and therefore
allow more accurate modeling of this very luminous source.

Thus far the lensing galaxy for Q1208+1011 has not been directly
imaged, consistent with the expectation that it is 4-6 magnitudes
fainter than the quasar (Bahcall et al. 1992, Kochanek 1991).  The
small separation indicates that a galaxy at relatively high redshift,
$z \ga 0.5$, is likely responsible for the lensing (Turner et
al. 1984).  For this system and others with suspected high-redshift
lensing galaxies, it may be possible and even necessary to identify the
lens by its {\em absorption} properties, rather than by its emission.
With few exceptions, galaxies within $\sim 30 h^{-1}$\,kpc of the
quasar line of site cause MgII or CIV absorption (Steidel 1997;
Steidel 1993), so one would expect an metal-line absorption system at
the lens galaxy redshift.

For Q1208+1011, the only published analysis of possible lens redshifts
based on absorption lines has been by Magain et al. (1992).  They
re-analyzed the absorption line data presented by Steidel (1990) and
suggested at least 18 possible metal-line absorption systems, spanning
redshifts from 0.3741 to 2.9157, with the majority in the range $2.5 <
z < 3.1$.  They proposed that the most likely lens system was at
redshift 2.9157 and derived a corresponding mass estimate for the
lens. However, lens models indicate that such a high redshift location
is highly unlikely (see Section~\ref{sec:lens_model}).  Furthermore,
most of low redshift identifications were based on just two doublet lines
within the Ly-$\alpha$ forest, whereas Bechtold and Yee (1995) have
shown that the false detection rate for doublets in the forest is
quite high (see also Section~\ref{sec:abs_lines}).

To constrain the lens redshift more reliably in Q1208+1011,
we obtained a high-resolution spectrum in March 1996 with the
echellette spectrograph on the Multiple Mirror Telescope (MMT).  In
this paper we examine the new and published spectra and provide an
updated list of high-confidence metal-line absorption systems
(Section~\ref{sec:abs_lines} and Section~\ref{sec:lens_model}).  We
then combine this with a simple model of the gravitational lens system
to constrain the possible lens redshifts
(Section~\ref{sec:lens_model}).  We show that the high-redshift ($z >
2.5$) and low-redshift ($z < 0.4$) candidates proposed by Magain et
al. can be ruled out with high confidence, and that the current
spectra effectively probe about 40\% of the redshift range in which
the lens is expected.  In that range, there is only one known
metal-line absorption system, an MgII absorber at $z=1.1349$.  In
Section~\ref{sec:discussion} we consider the possibility that this
system is the lensing galaxy and discuss the implied parameters of the
galaxy. We also calculate the expected galaxy IR luminosity.

\section{Observations and Data Reduction}
\label{sec:abs_lines}

The spectrum shown in Figure~\ref{fig:spectrum} represents a total of
6.7\,hours of integration on Q1208+1011, taken in eight 3000~second
exposures on 1996 March 26.  The spectrum has been normalized to the
continuum and the dotted line below the spectrum shows the 1-$\sigma$
(per pixel) error array. These were obtained at the MMT with the blue
channel echellette, in which the MMT blue channel spectrograph is used
with a quartz prism cross-disperser.  The echellette grating has 240
l/mm and is used in orders 7-17, giving coverage from 3100-8150\AA.
The slit was 10\arcsec$\times$1\arcsec\ long, giving spectral
resolution of $\sim 45-50$~km~s$^{-1}$.  The CCD detector was a $3072
\times 1024$ Loral CCD.  The usable range of our spectrum is 4000 --
6360\AA.

The observations included bias frames, quartz flats, and
copper-helium-neon-argon comparison lamps bracketing each exposure.
The spectra were reduced using the IRAF {\tt noao.imred.echelle}
package and the procedure described in Hamuy and Wells (1989).  After
wavelength calibration, the individual echelle orders were
approximately flux calibrated using a standard star which was observed
and reduced in the same manner.  The variance array for each spectrum
was calculated by the IRAF spectral extraction tool (\texttt{apall}),
using the known noise characteristics of the CCD.  The eight exposures
were combined with a weighted average based on average
signal-to-noise.  The normalization continuum used in
Figure~\ref{fig:spectrum} was determined by the iterative fitting
method described in Aldcroft et al. (1994).

\section{Metal-Line Absorption Systems}

The radio-quiet quasar Q1208+1011 (Hazard et al. 1986, Sargent et al. 1986) was
discovered in an objective prism survey.  This object is too faint to have been
observed with IUE in the UV, and all existing HST spectra are at $\lambda \gs
3700$\AA (Bahcall et al. 1992).  The original discovery spectrum (Sargent et
al. 1986) covers 3200 -- 10000\AA\ ($\lambda_{rest}=$ 665 -- 2078\AA) with
detectable flux throughout, showing that there are no optically thick
absorption clouds in the range 2.5$\ls z \ls$3.8.

The moderate resolution (100\kms) spectra of Steidel (1990) and Bechtold
(1994) spectra cover 5350 -- 7695\AA\ and 5470 -- 6120\AA,
respectively. Those authors identified three high-redshift CIV absorbers
above the \lya\ forest at $z = 2.9137, 2.8606, 2.8573.$ Steidel (1990)
also proposed a fourth system at $z= 3.1985$, however Magain et
al. (1992)  argued that it is improbable because the wavelength match
is very poor.
All systems are optically thin at the Lyman limit ($\tau_{total} <1$).
Both Steidel and Bechtold identified the line at 5970\AA\ as C~IV
1548\AA\ at $z$=2.8573 and the line at 5985\AA\ as C IV 1550\AA\ at
$z$=2.8593 blended with C~IV 1548\AA\ at $z$=2.8606.  This is
supported by the existence of a weak line with rest equivalent width =
$0.35\pm 0.04$\AA\ ($\lambda = 4687.7\pm0.1$\AA) in the \lya\ forest
at the correct redshift.  However, our new higher resolution spectrum
($\sim 45$\kms) suggests a much more plausible set of identifications
for the complex between 5970\AA\ and 5990\AA, as shown in Table~1 and
Figure~\ref{fig:spectrum_detail}.  We now see that the lines at
5970\AA\ and 5985\AA\ (Figures~\ref{fig:spectrum_detail} and
\ref{fig:mgii}) have very similar profiles and an exact wavelength
match for Mg II 2896\AA\ and 2803\AA\ at $z=1.1349$.  Likewise, the
wavelength match and the profiles of the lines at 5980\AA\ and
5990\AA\ imply an identification of CIV at $z=2.8636$.  The
subcomponent structure for these features was not visible in either
the Steidel or Bechtold spectra.  The weak
\lya\ at $z = 2.8606$ mentioned previously is therefore not associated with a
metal-line system.

Adopting the identification of the $z=1.1349$ Mg II system, our spectra cover
several other potential absorption lines, as shown in Figure~\ref{fig:mgii}.
{}From the figure it appears that the Mg\,II system has two components, at $z =
1.1346$ and $z = 1.1351$.  Redward of the \lya\ forest, the MgI 2852\AA\ line
is not detected to a 3-$\sigma$ rest equivalent width limit of 0.09\AA.  Our
spectrum also covers the Fe II lines in the range $\lambda\lambda 2344 -
2600$\AA, although all lie in the Lyman alpha forest. We detect a line at the
right position to be Fe II 2600\AA\ ($z = 1.1351$), although it is probably
blended with weak Lyman alpha absorption.  There is also a hint of a line
$z=1.1346$. The others lines (2586, 2382, 2374, 2344\AA) appear to be blended
with strong Lyman alpha features, as seen for Fe\,II 2382 in
Figure~\ref{fig:mgii}.

Magain et al. (1992) reanalyzed the Steidel (1990) spectrum of
Q1208+1011 and suggested at least 18 possible absorption systems,
spanning redshifts from 0.3741 to 2.9157, with the majority in the range
$2.5 < z < 3.1$.  These identifications were based primarily on doublets
identified in the Lyman alpha forest region, and while any given system
is hard to rule out, the number of spurious detections of doublets in
the forest by chance is actually quite high.  Bechtold and Yee (1995)
used simulated pure \lya\ forest spectra with the line density
appropriate to $z\sim 3-4$ and searched for spurious metal-line systems
with physically allowable line ratios.  They found that to reject
chance matches requires a minimum of 4 lines matching in central
wavelength. Most of the systems suggested by Magain et al in the forest
are based on only 2 lines, and are hence quite likely to be spurious.

Nevertheless, we examined in detail three of the systems ($z = 0.3733,
0.3843,$ and 0.9274)\footnote{
The actual redshifts reported by Magain et al, based on Steidel
(1990), were $z=0.3741, 0.3850,$ and 0.9281, respectively.  However,
all the lines in the Q1208+1011 line list from Steidel appear
blueshifted by about 150\kms\ with respect to the lines in our
spectrum {\em and} those in Bechtold (1994).  We believe the Steidel
line list to be in error and we refer to these systems by the
redshifts derived from our new spectrum.}
suggested by Magain et al. because their redshifts make them relevant to
the possible identification of the lens.  In Figure~\ref{fig:magain} we
plot the key doublet for each system, along with a third strong line
which could lend credibility to the identification.  The systems at
$z=0.3733$ and 0.3843 were suggested based on supposed Ca II H\&K
doublets.  Our spectra confirm the lines at the right wavelengths for
these identifications, and but we fail to detect Mg II 2796\AA, 2803\AA\
for either system in our spectra.  These systems are therefore likely
spurious.  The possible Mg\,II absorber at $z=0.9274$ appears to have a
small wavelength mismatch at higher resolution, and also shows no
evidence for Fe\,II 2382 nor Fe\,II 2600 absorption.  This system is
also likely to be spurious.

Finally, we searched our spectrum for metal systems in the Lyman alpha
forest using our code described by Bechtold and Yee (1995) and
requiring at least five matching lines with physically possible line
ratios.  We found no plausible systems.

A graphical summary of the metal-line absorption systems in
Q1208+1011, and the redshifts which have been searched, is given in
Figure~\ref{fig:absorbers}.  The solid lines show regions that have
been reliably searched for absorption. For Ca\,II, Mg\,II and C\,IV
doublets, this implies the region above the
\lya\ forest.  The dashed lines show the redshift ranges within the
\lya\ forest in which we searched for metal systems with at least four
matching lines.  In these ranges only very strong absorbers would be
detected.  The crosses indicate detected absorption systems.  It
should be noted that Ca\,II is generally weaker and much less common
than Mg\,II, and the absence of Ca\,II for $0.48 < z < 0.95$ does not
imply that there are no galaxies in the quasar line-of-sight in this
redshift range.

We conclude that there are five high-confidence metal-line absorption
systems at redshifts z = 1.1349, 2.8626, 2.9118, 2.9136, 2.9149
present in the spectrum of Q1208+1101. Now, we consider the
possibility for each of them being the location of the lensing galaxy.

\section{The Lens}
\label{sec:lens}
\subsection{Constraints on the lens redshift}
\label{sec:lens_model}

The redshift of the lensing galaxy can be identified through its own
emission or absorption of the background quasar. For higher
redshift lenses, searching the metal-line absorption systems is the more
appropriate method, especially if the lensing galaxy has not been directly
imaged.  The strategy of searching for the lensing galaxy redshift
by means of its absorption has the advantage that there is no flux
bias in the lens identification, but there are also limitations:
\begin{itemize}
\item The absorption doublet must occur outside the quasar Ly-$\alpha$
forest.  For a quasar redshift of 3.815, this means that galaxy
absorption can only be reliably detected for $z_{abs} > 1.09$ (MgII)
or $z_{abs} > 2.76$ (CIV);

\item For ground based observations, the lens redshift must have $z \ga
0.18$ to detect the MgII doublet.

\item The line-of-sight to a typical quasar intersects on average $\sim
1$ Mg\,II absorber per unit redshift (Steidel 1992), and $\sim 2$
C\,IV absorbers per unit redshift (Sargent et al 1988).  Additional
information or constraints are therefore needed to narrow the search.
\end{itemize}

One key constraint is the probability distribution for the lens
redshift based on gravitational lens theory.  This helps in
understanding what range of redshifts is plausible, and allows us to
compute the probability that the lens redshift is below the Mg~II
detection threshold (z=1.09).  To do this, we use the standard model
of gravitational lens statistics which combines a singular isothermal
sphere (SIS) lens model with a Faber-Jackson or Tully-Fisher relation
and a Schechter luminosity function (e.g.\ Kochanek 1996).  We use
this model to compute the probability density $dP/dz$ for a lens at
redshift $z$ producing the observed image separation (see Kochanek
1992).  Using a Schechter luminosity function to represent the galaxy
population at high redshift may be questionable given
evidence for population evolution (e.g.\ Kauffmann, Charlot \& White
1996; Zepf 1997).  However, preliminary studies suggest that
gravitational lens statistics are not dramatically affected by
evolutionary effects (Mao \& Kochanek 1994; Rix et al.\ 1994), and
since a full treatment of lens statistics with evolution is not yet
available we adopt the standard approach with no population evolution.

Figure~\ref{fig:prob_dist} shows the cumulative probability $P(<z)$
that the lens galaxy in Q1208+1011 is at a redshift less than $z$,
computed for early-type galaxies (ellipticals and S0s) and spiral
galaxies in two different cosmologies.  The median redshift and 90\%
confidence interval are marked for each model, and the Mg~II detection
threshold at $z=1.09$ is indicated.

Of the five metal-line absorption systems (see Table 1) we detect, the
four high-redshift CIV systems are outside the 99\% confidence
interval and thus are unlikely to be associated with the lens galaxy.
By contrast, it is quite plausible that the Mg\,II system at
$z=1.1349$ could be associated with the lens galaxy, because the
median expected redshift is 1.34--1.56 for an early-type galaxy and
1.05--1.22 for a spiral galaxy.  Combining the information in
Figures~\ref{fig:absorbers} and
\ref{fig:prob_dist}, we find that the redshift range which has been
searched for absorbers covers 39-41\% (early-type lenses) and 34-40\%
(spiral lenses) of the lens probability distribution.  Therefore the
$z=1.1349$ system is a good lens candidate, but it is by no means secure.

It is also worth nothing that there is less than an 8\% probability that
the lens in Q1208+1011 has $z < 0.4$.  Thus even if one of the two Ca\,II
systems suggested by Magain et al (1992) were real, it is unlikely that
it could be the gravitational lens.

\subsection{Lensing Galaxy Properties}
\label{sec:discussion}

Because the separation between the two quasar images is only 0.\arcsec 47,
the mass of a normal galaxy is adequate to produce the lensed images.
Assuming the Singular Isothermal Sphere (SIS) model for the lensing galaxy
we can estimate the velocity dispersion and the enclosed mass for a given
redshift.  In the redshift range which we have searched, there is only one
candidate lens redshift, the $z=1.1349$ MgII system.  However, since there
is a significant probability that this is not the lens redshift, we also
calculate the parameters for several other interesting redshifts: $z=0.4$
low-redshift case (low end of the 90\% probability range); $z=2.4$
high-redshift case (high end of the 90\% probability range); and $z=2.9$
C\,IV case, corresponding to the known C\,IV absorption systems. In all the
calculations below, we assume $\rm \Omega_0=0.1$ and $\rm
H_0=100h~km~sec^{-1} Mpc^{-1}$.

The mass of the galaxy can be obtained from:

$$ M  \sim {4 \theta^2  \over 9} \, {D_l D_s \over D_{ls}}$$

\noindent
where $\theta$ is the image separation, M is mass of the lens,
$D_l, D_s, D_{ls}$ are the angular diameter distances to the lens, to
the quasar and between the lens and the quasar respectively (see
the review by Blandford \& Narayan, 1992).  The corresponding velocity
dispersion is related to the image separation by:

$$ \theta = 4 \pi {\sigma^2 \over c ^2} \, {D_{ls} \over D_s} = 2.6 \arcsec
\sigma ^2 _{300} {D_{ls} \over D_s} $$

\noindent
where $\sigma = 300 \times \sigma _{300} $~km~s$^{-1} $ is a velocity
dispersion.

Table 2 contains the calculated mass and velocity dispersions for the
four considered lens redshifts.  The main uncertainty on the mass is
related to cosmology and the uncertainty on the Hubble constant.

The required mass ($\sim 2.8 \times 10^{11} M_{\odot}$) and the velocity
dispersion ( $\sim 202$~km~s$^{-1}$ for the lens at $z=1.1349$ are quite
typical of normal galaxies.  We believe the Mg II absorption system at
z=1.1349 is a strong candidate to be the lensing galaxy.  Absorption of
the kind and strength we see is, with few exceptions, associated with a
galaxy within $\sim30 h^{-1}$\,kpc of the line of sight (Le Brun, et al. 1995;
Steidel, 1993). This implies that there is a galaxy within $\sim
4$\arcsec\ of the Q1208+1011 pair.  Additionally, only in rare cases is
there a galaxy within $\sim 30h^{-1}$\,kpc which {\em does not} cause Mg\,II
absorption (Steidel 1993).

Thus far the lensing galaxy has not been detected -- pre-Costar HST
imaging with the PC (Bahcall et al. 1992) limits the galaxy to have
V$> 20.7$ if it is more than 0.\arcsec 5 from the brighter component,
or V$>19$ if the galaxy is between the two images.

Given the mass estimate for this system, we can predict its brightness
using the Tully-Fisher and Faber-Jackson relations.
If the galaxy is a disk system, the velocity dispersion implies that
it is about 1 magnitude brighter than $L^*$; if it is an elliptical,
it is 0.3 magnitudes fainter than $L^*$.  Figure 5 shows predicted
lens galaxy magnitudes in HST $V$ (F555W) and $H$ (F160W) bands.  The
luminosities were estimated by combining an SIS lens model with the
Faber-Jackson or Tully-Fisher relation, and the magnitudes were then
estimated by applying $K$ and evolutionary corrections computed from
the spectral evolution models of Bruzual \& Charlot (1993).  (See
Keeton, Kochanek \& Falco 1997 for details.)
The predicted apparent magnitude in the visible is V=24.1-25.4 (see
Figure~\ref{fig:prob_dist}), much fainter than the limit $V\sim20.7$
placed by Bahcall et al (1992) from the pre-Costar PC on HST.  The
predicted near-IR magnitude is $H\approx19.2-20.6$ (see
Figure~\ref{fig:prob_dist}), or $K\approx20.2-21.6$.  This is near the
faint end of the range of luminosities of galaxies selected by the
presence of Mg\,II absorption and described by Steidel \& Dickinson
(1995).  They presented data which showed that, for 5 Mg\,II systems
with $1.0 < z < 1.2$, the galaxy causing the absorption had K
magnitude between 18.5 and 20.0.  We have simulated NICMOS
observations to determine whether such a galaxy will be easily
visible.  We assume the galaxy is centered between the quasar images,
synthesize a test image, and remove the quasar images using a
synthesized point-spread-function.  We find that a four-orbit exposure
with the low background H-band (F160W) filter might give a detection
with sufficient signal to estimate a lens model and the corresponding
magnification.  A single-orbit exposure such as the one planned for
Cycle~7 (Falco et al.)  \footnote{Preliminary NICMOS images of
Q1208+1011 have recently become available on the CASTLE Web page:
{\texttt http://cfa-www.harvard.edu/glensdata/1208.html}. The galaxy
is not apparent in the image consistent with the predicted magnitude.}
migth detect the core of the galaxy but will not likely trace the
profile very far beyond the quasar image.
We note that the image separation of 0.$\arcsec$47 corresponds to
$\sim 3h^{-1}$~kpc, slightly smaller than typical scale lengths and
effective radii of $L^*$ galaxies.

\bigskip

We have combined our high-resolution spectra of the metal-line
absorption systems towards the lensed quasar Q1208+1101 with
gravitational lensing models.  We find the MgII absorber at
z=1.1349 to be a plausible candidate for the lensing galaxy.

\acknowledgments{We would like to thank Christopher Kochanek for
very helpful discussion. AS and TLA were supported by NASA Contract NAS
8-39073 (ASC).}

\newpage
\centerline{REFERENCES}

Aldcroft, T., Bechtold, J. and Elvis, M., 1994, ApJ, 93, 1

Antonucci, R., 1994, in: ``Multi-wavelength continuum emission of
AGN'', IAU Symposium No. 159., Ed. Courvoisier \& Blecha, Kluwer,
Dordrecht, p.301

Bahcall, J.N. et al 1992, Ap.J. 392, L1

Bechtold, J. 1994 ApJ Supp 91,1

Bechtold, J. and Yee, H.K.C. 1995 AJ, 110, 1984.

Blandford, R. \& Narayan, R. 1992, ARAA

Bruzual, G., \& Charlot, S. 1993, ApJ, 405, 538

Czerny, B., 1994, in:``Multi-wavelength continuum emission of
AGN'', IAU Symposium No. 159., Ed. Courvoisier \& Blecha, Kluwer,
Dordrecht, p.261

Fontana et al. 1997, in Proceedings of the 13$^{th}$ IAP Colloquium:
Structure and Evolution o the Intergalactic Medium from QSO Absorption
Line Systems help in Paris on July 1997, Eds. P.Petitjean and S.Charlot,.

Hazard, C., McMahon, R.G., Sargent, W.L.W., 1986, Nature, 322, 38

Kauffmann, G., Charlot, G., \& White, S. D. M. 1996, MNRAS,
283, L117

Keeton, C.R., Kochanek, C.S., Falco, E.E., 1997 Ap.J. submitted
(astro-ph/9708161)

Keeton, C.R., \& Kochanek, C.S., 1996, in Astrophysical Applications
of Gravitational Lensing, ed. C.S.Kochanek \& Hewitt, J.N., Dordrecht:
Kluwer, 419

Kochanek, C.S., 1991, Ap.J., 373, 354

Kochanek, C. 1992, Ap.J., 384, 1

Kochanek, C.S. 1996, ApJ, 466, 638

Le Brun, V., et al. 1995, A\&A, 279, 33

Narayan, R., \& Wallington, S. 1992, in Gavitational Lens, Proceedings of
the IAU Symposium No 130. Eds.  Kayser, R., Schramm, T., Nieser, L. Springer
Verlag 1992

Magain, P., Surdej, J., Vanderriest, C., B. Pirenne, D. Hutsemekers,
1992, A\&A 253, L13.

Mao, S. D., \& Kochanek, C. S. 1994, MNRAS, 268, 569

Maoz, D. et al. 1992, Ap.J., 386, L1

Rix, H.-W., Maoz, D., Turner, E. L., \& Fukugita, M. 1994,
ApJ, 435, 49

Sargent, W.L.W et al., 1986, Nature, 322, 40

Siemiginowska, A., Bechtold, J., Tran, K.-V., Dobrzycki, A., 1996, in
Science with the Hubble Space Telescope-II, Proceeding of a Workshop
held in Paris, December 4-8, 1995, Eds. Benvenuti,P., Macchetto, F.D.,
Schreier, E.J., p.181.

Steidel, C.C., Dickinson, M., Meyer, D.M., Adelberger, K.L., Sembach,
K.R., 1997, Ap.J., 480, 568.

Steidel, C. C., in  Environment and Evolution of Galaxies, Eds. Shull, M.,
Thronson, H., Kluwer 1993.

Steidel, C. C. 1990 ApJ Supplement Ser 72,1

Steidel, C. C., \& Dickinson, M., Preprint, 1995

Turner, E.L., Ostreiker, J.P., Gott III, J.R., 1984, Ap.J. 284, 1.

Zepf, S. E. 1997, Nat, 390, 377
\newpage

\figcaption[fig1.ps]{Spectrum of Q1208+1011 obtained with the MMT Blue
Spectrograph and cross-dispersed echellette grating.  The spectrum has
been divided by a low order fit to the continuum \label{fig1}
\label{fig:spectrum}}

\figcaption[fig2.ps]{ Portion of the spectrum shown in Figure 1 containing the
Mg II and C IV systems discussed in the text.  All wavelengths
are vacuum values.  \label{fig2} \label{fig:spectrum_detail}}

\figcaption[fig3.ps]{Regions near the indicated lines for the $z=1.1349$
Mg\,II system as a function of velocity, where the indicated redshifted
line centers have been used to define zero velocity. \label{fig:mgii}}

\figcaption[fig4.ps]{
Regions near the indicated lines for three systems suggested by Magain et
al (1992).  Each is plotted as a function of velocity, where the
indicated redshifts have been used to define zero velocity.
\label{fig:magain}}

\figcaption[fig5.ps]{
Spectral coverage for Q1208+1011 for Lyman limit, C\,IV, and Mg\,II, as a
function of redshift.  Solid line regions have been reliably searched for
absorption and dashed lines show ranges within the \lya\ forest.
The crosses mark detected systems.
\label{fig:absorbers}}

\figcaption[fig6.ps]{ Top: Cumulative probability $P(<z)$ that the lens
galaxy has redshift less than $z$, computed for an early-type lens
galaxy (left) and a spiral lens galaxy (right) in two different
cosmologies.  The filled and open boxes indicate the median
redshift and 90\% confidence interval.  The vertical dashed
line indicates the Mg~II detection threshold at $z=1.09$.
Bottom: Predicted lens galaxy magnitudes in HST $V$ (F555W)
and $H$ (F160W) bands, again for an early-type lens galaxy
and a spiral lens galaxy in two cosmologies.
\label{fig:prob_dist}}

\begin{figure}[ht]
\centering
\plotone{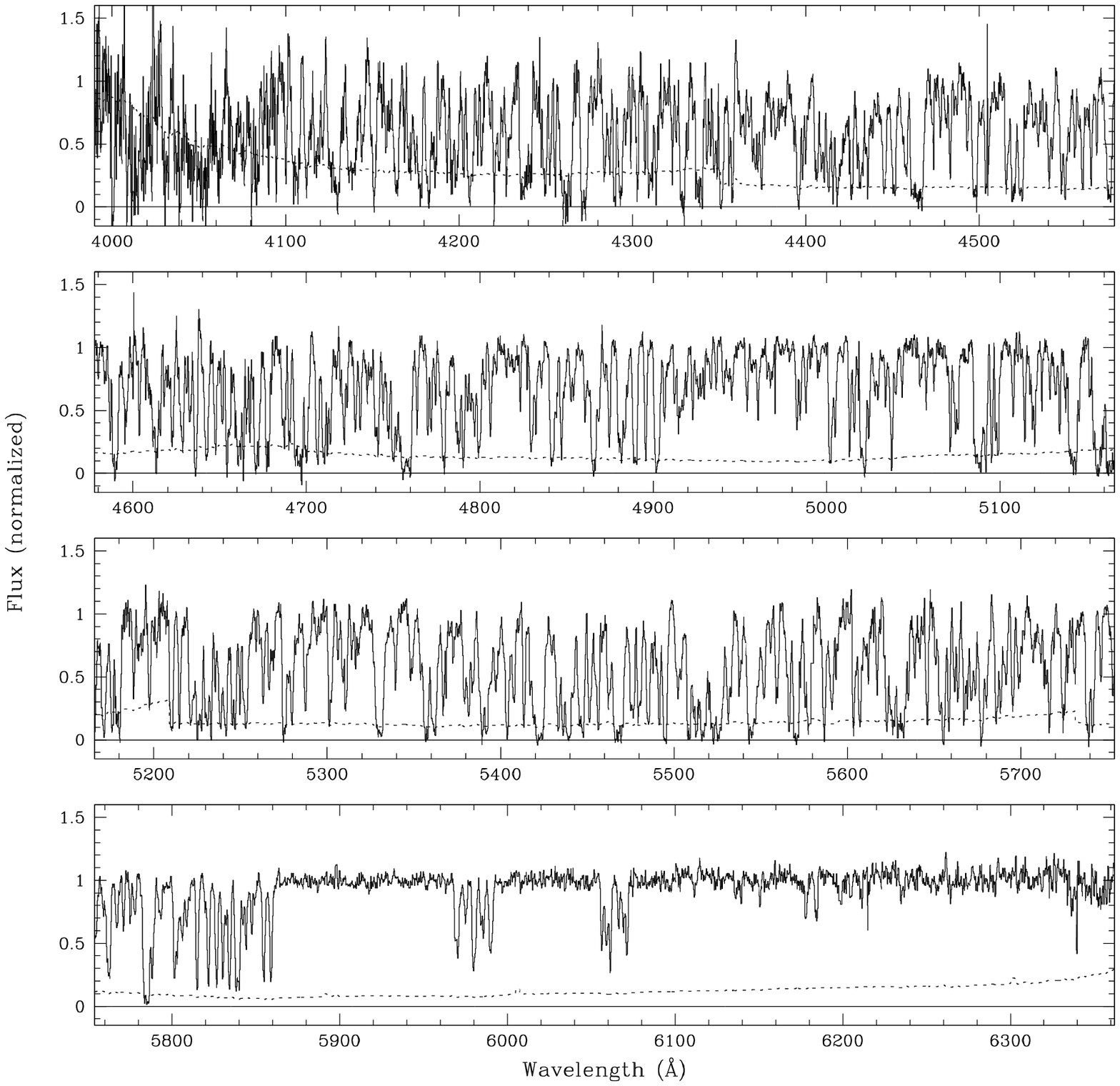}
\end{figure}

\begin{figure}[ht]
\centering
\plotone{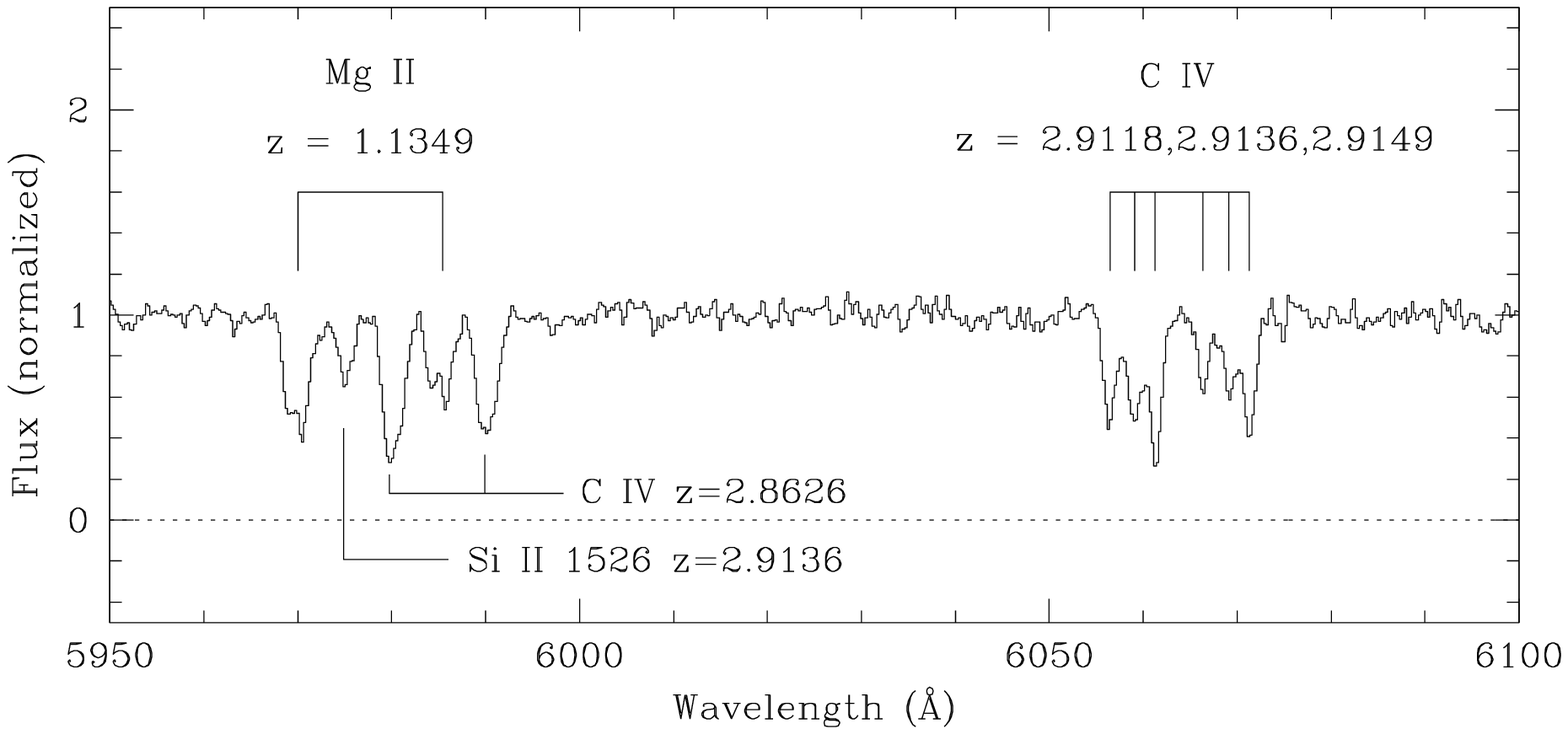}
\end{figure}

\begin{figure}[ht]
\centering
\plotone{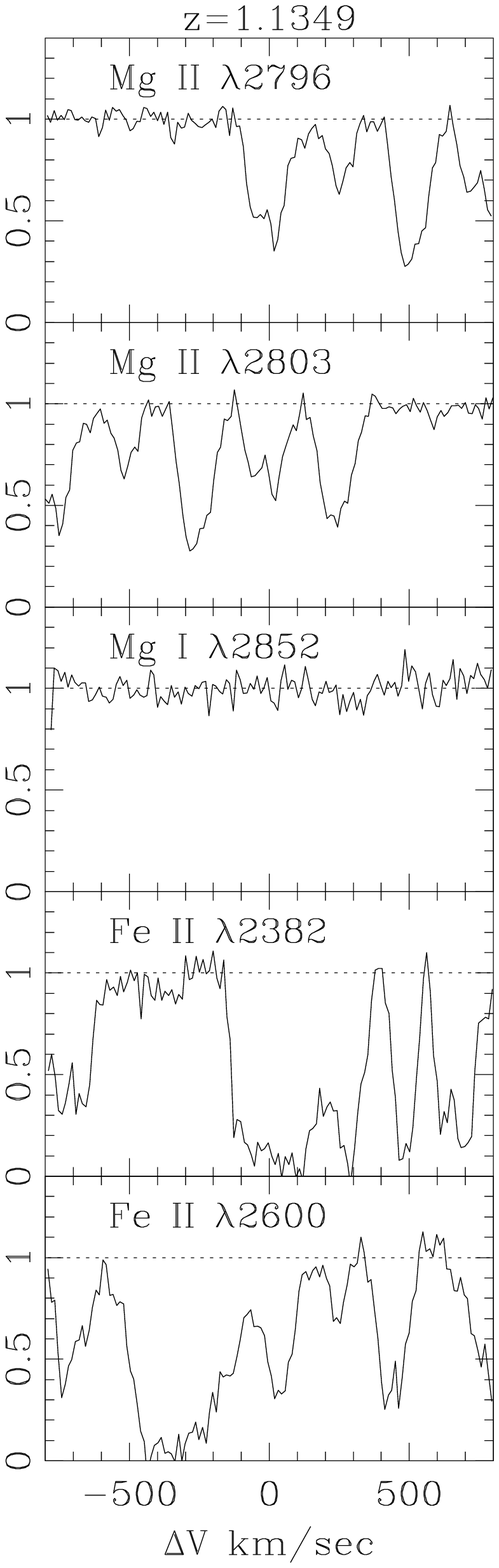}
\end{figure}

\begin{figure}[ht]
\centering
\plotone{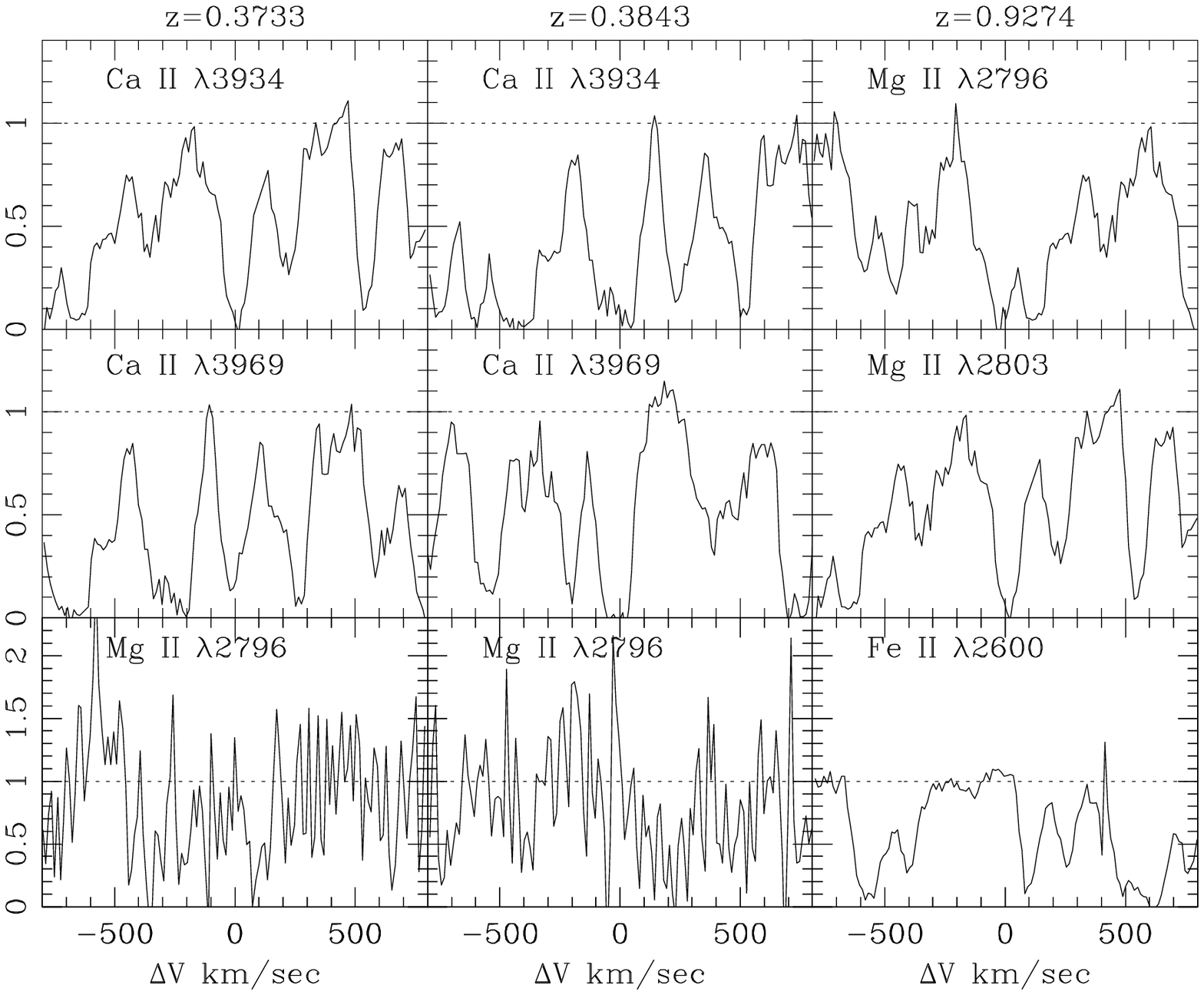}
\end{figure}

\begin{figure}[ht]
\centering
\plotone{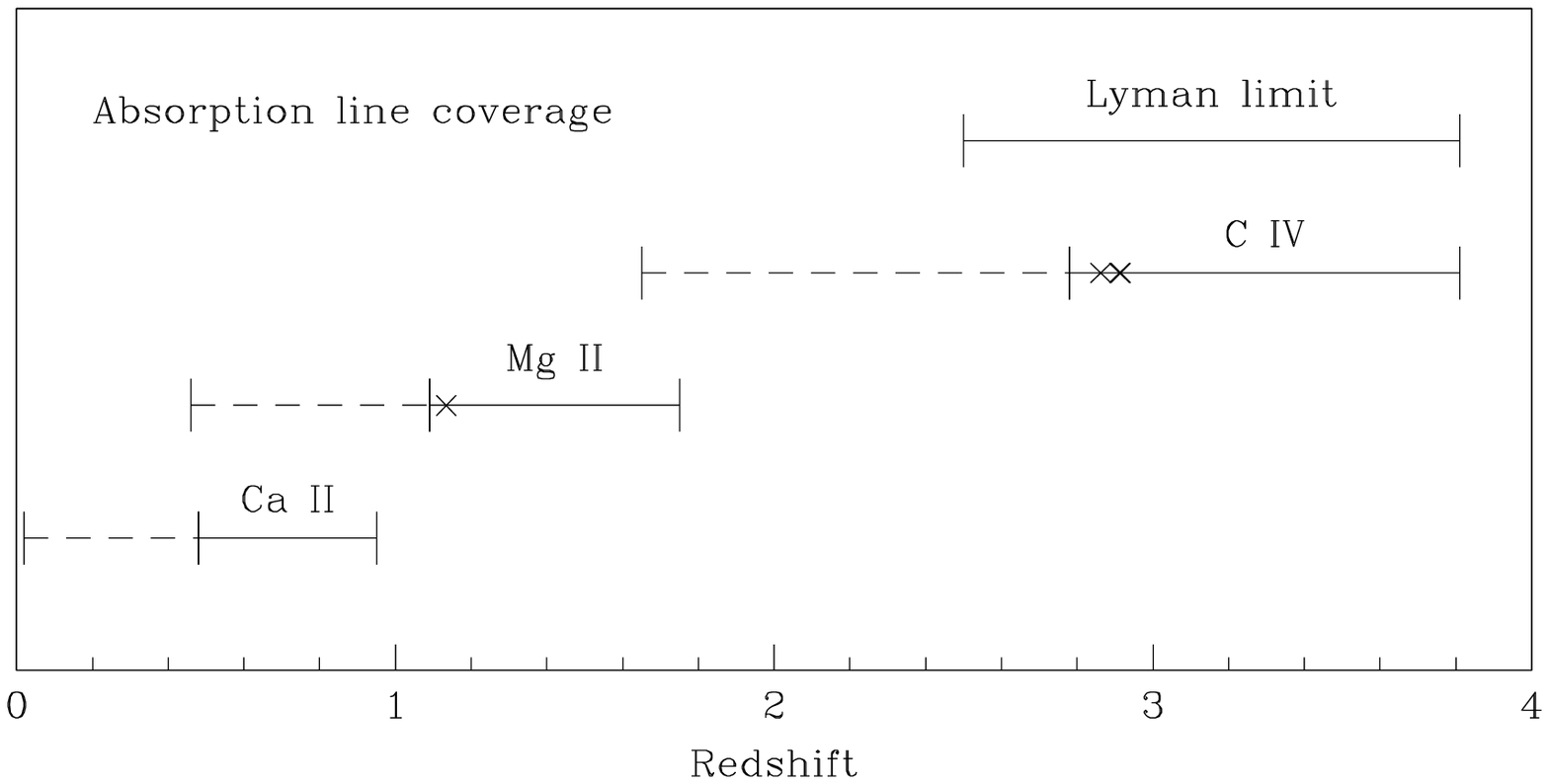}
\end{figure}

\begin{figure}[ht]
\centering
\plotone{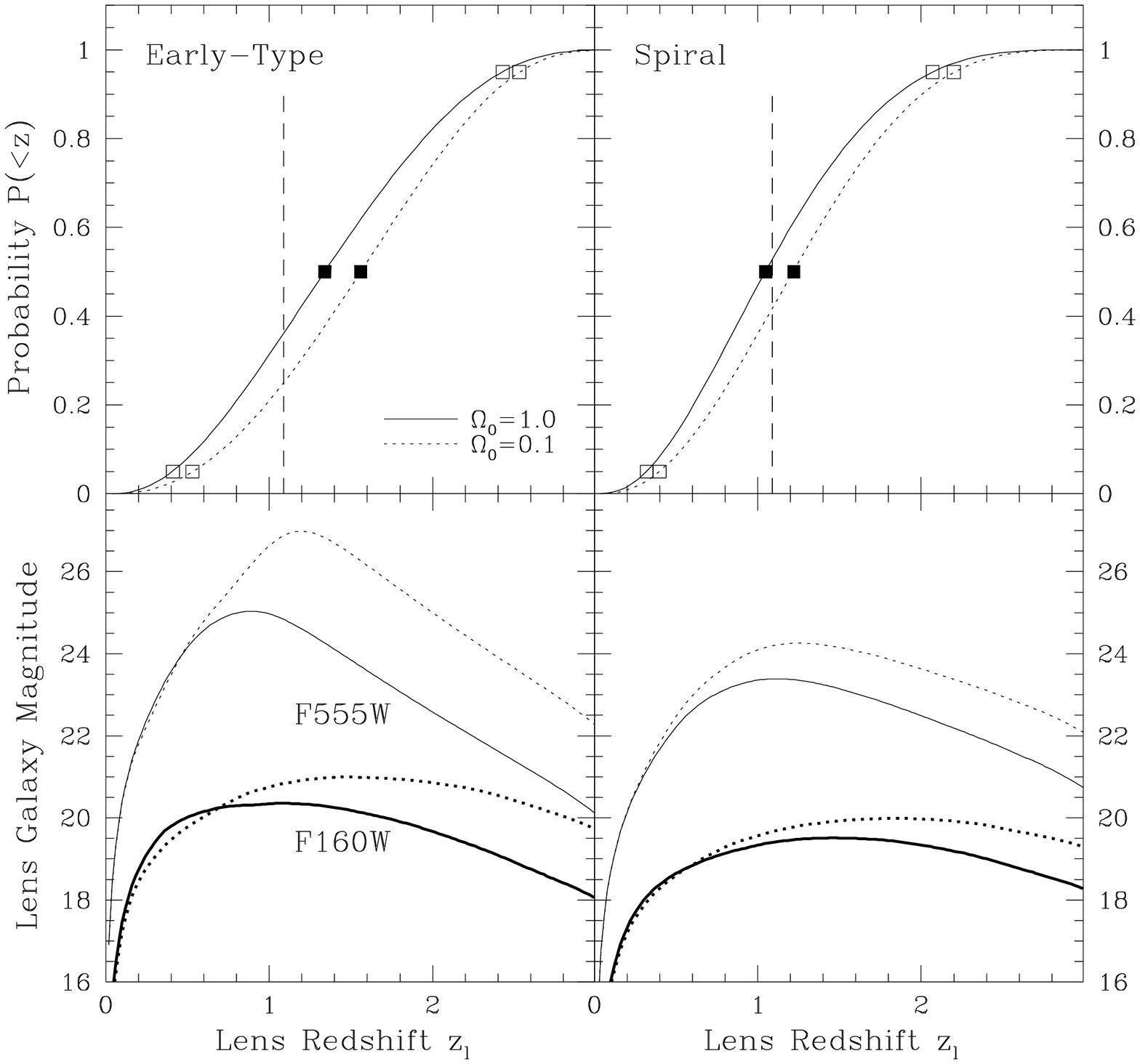}
\end{figure}

\newpage

\begin{table}
\begin{center}
\caption{Absorption Lines}
\bigskip
\begin{tabular}{cccccccc}
\tableline
 $\lambda _{obs} $ &  $\sigma (\lambda)$ &  $W_{obs}$(\AA ) & $\sigma (W)$ &
SL$^a$ &  FWHM (\AA ) & Line ID   & $z_{abs}$ \\
\tableline
&&&&&&&\\
 5969.98 &  0.07 &  1.67& 0.08 & 21.2 & 2.65 & Mg II(2796) & 1.1349 \\
 5974.92 & 0.11  & 0.61 &0.07  & 8.8  &1.79  &Si II (1526) & 2.9136 \\
 5980.05 & 0.06  & 1.82 &0.08  &22.6  &2.72  &C IV (1548)  & 2.8626 \\
 5985.10 & 0.10  & 1.16 &0.08  &14.1  &2.83  &Mg II (2803) & 1.1348 \\
 5990.02 & 0.07  & 1.52 &0.08  &18.8  &2.68  &C IV (1550)  & 2.8626\\
 6056.40 & 0.08  & 0.86 &0.08  &10.4  &1.60  &C IV (1548)  & 2.9119\\
 6058.99 & 0.06  & 0.97 &0.08  &12.7  &2.04  &C IV (1548)  & 2.9135 \\
 6061.20 & 0.04  & 1.08 &0.07  &15.1  &1.30  &C IV (1548)  & 2.9149\\
 6066.27 & 0.10  & 0.51 &0.07  & 6.8  &1.37  &C IV (1550)  & 2.9118\\
 6069.01 & 0.09  & 0.68 &0.08  & 8.5  &1.64  &C IV (1550)  & 2.9135\\
 6071.19 & 0.05  & 0.91 &0.07  &12.5  &1.40  &C IV (1550)  & 2.9149\\
&&& & & & &  \\
\end{tabular}
\end{center}
\tablenotetext{a} {significance level as in Aldcroft et al. (1994)}
\end{table}

\begin{table}
\caption{Lens Parameters}
\bigskip
\begin{tabular}{cccc}
\tableline
redshift &    mass &  	$\sigma$  \\
z$_l$ & 10$^{11} h^{-1}$ M$_{\odot}$ &  km~s$^{-1}$ \\
\tableline
&&&\\
0.4 & 1.05 & 154 \\
1.1349 & 2.8  & 202 \\
2.4 & 8.1& 325 \\
2.91   & 13.8 &  425 \\
&&&\\
\tableline
\end{tabular}
\end{table}

\end{document}